\documentclass{aa}
\usepackage{graphics}
\usepackage{amsfonts}
\input{epsf}
\newcommand{\bfo}[1]{\mbox{\boldmath $#1$}}
\def\bvarphi{\mbox{\boldmath $\varphi$}}

\begin{document}
\newcommand{\beq}{\begin{equation}}
\newcommand{\eeq}{\end{equation}}
\def\change{\sffamily}
\def\la{\hbox{\raise.35ex\rlap{$<$}\lower.6ex\hbox{$\sim$}\ }}
\def\ga{\hbox{\raise.35ex\rlap{$>$}\lower.6ex\hbox{$\sim$}\ }}
\def\runit{\hat {\bf  r}}
\def\phunit{\hat {\bfo \bvarphi}}
\def\etaunit{\hat {\bfo \eta}}
\def\zunit{\hat {\bf z}}
\def\zetaunit{\hat {\bfo \zeta}}
\def\xiunit{\hat {\bfo \xi}}
\def\beq{\begin{equation}}
\def\eeq{\end{equation}}
\def\beqa{\begin{eqnarray}}
\def\eeqa{\end{eqnarray}}
\def\sub#1{_{_{#1}}}
\def\order#1{{\cal O}\left({#1}\right)}
\newcommand{\sfrac}[2]{\small \mbox{$\frac{#1}{#2}$}}
%
%
\title{{On the viability of the shearing box approximation for numerical
       studies of MHD turbulence in accretion disks}}

\author{O. Regev\inst{1,2} \and O.M. Umurhan\inst{1,3,4}}

   \offprints{O. Regev \ \email{regev@astro.columbia.edu}}

   \institute{Department of Physics, Technion-Israel Institute of
    Technology, 32000 Haifa, Israel\
\and
     Department of Astronomy, Columbia University,
     New York, NY 10025, USA
\and
     Department of Geophysics and Space Sciences, Tel-Aviv University,
     Tel-Aviv, Israel
\and
     Department of Astronomy, City College of SF,
      San Francisco, CA 94112, USA
}

\date{}


  \abstract
     {Most of our knowledge of the nonlinear development of the magnetorotational
      instability (MRI) relies on the results of numerical simulations employing
      the shearing box (SB) approximation. A number of
      difficulties arising from this approach have recently been pointed out
      in the literature.}
     {We thoroughly examine the effects of the assumptions made and numerical
       techniques employed in SB simulations. This is done to
      clarify and gain a better understanding of those difficulties, in addition
      to a number
      of additional serious problems raised here for the first time,
      and of their impact on the results.}
     {We used analytical derivations and estimates as well as comparative analysis to methods
     used in the numerical study of turbulence. We performed numerical experiments
     to support some of our claims and conjectures.}
     { The following problems, arising from the
      (virtually exclusive) use of SB simulations as a tool for the understanding
      and quantification of the nonlinear MRI development in disks, are analyzed
      and discussed:
      (i) inconsistencies in the application of the SB
      approximation itself;
     (ii) the limited spatial scale of the SB;
     (iii) the lack of convergence of most ideal mgnetohydrodynamical (MHD) simulations;
     (iv) side-effects of the SB symmetry and
      the non-trivial nature of the linear MRI; and
    (v) physical artifacts arising from the very small box scale due to periodic boundary conditions.}
    {The computational
     and theoretical challenge posed by the MHD turbulence problem in accretion disks cannot
     be met by the SB approximation, as it has been used to date. A new strategy
     to confront this challenge is proposed, based on techniques widely used
     in numerical studies of turbulent flows - developing (e.g., with the help of local
     numerical studies) a sub-grid turbulence model and implementing it
     in global calculations.}

\titlerunning{Viability of shearing box approximation}

\keywords{accretion, accretion disks -- instabilities -- magnetohydrodynamics --
turbulence}

  \maketitle

\section{Introduction}
A physical mechanism for angular momentum transport outward is
needed for the theoretical modelling of accretion disks.
Prendergast \& Burbidge (1968) proposed the existence
of thin accretion disks to explain some galactic X-ray sources.
They realized that such disks must be characterized
by an enhanced viscosity, orders of magnitude larger that
the microscopic viscosity expected for the gas comprising the disks.
Given the likelihood that such flows are turbulent, they
estimated the value of this enhanced viscosity with the help
of a mixing-length theory.
A few years later, Shakura \& Sunayev (1973) and
Lynden-Bell \& Pringle (1974)
successfully bypassed the specific lack of understanding of turbulence
(a situation that is largely still with us) by employing
physically motivated parametrizations of this turbulent viscosity.
In doing so thin accretion disks, for the first time, could
be modeled in a variety of astrophysical settings. In turn, this
has brought about
significant progress in our understanding of these important systems.

Because the value and nature of the now famous nondimensional viscosity
parameter $\alpha$ still remains a theoretical unknown,
phenomenological approaches are usually employed to determine its value.
Such an approach also involves identifying an instability (or combination of
instabilities) that is responsible for the emergence of accretion disk turbulence.
This quest, according to this strategy, has become the focus of
intense (and sometimes contentious) research
for the three decades following the above mentioned ground-breaking studies.
Notwithstanding the current lack of a general theory of turbulence,
the identification of an instability process and, hopefully, also understanding
the transition from laminar flow into some turbulent-like steady
state (or at least quasi-steady one) has seemed to be the most promising strategy
to quantify angular momentum transport and energy generation
in accretion disks.
However, no definitive candidate hydrodynamical instability
had been convincingly identified and widely accepted before
Balbus \& Hawley (1991) discovered that weak magnetic fields
destabilize differential Keplerian rotation.
The linear magnetorotational instability (MRI; Velikhov 1959, Chandrasekhar 1960)
was shown to operate under conditions
characterizing rotationally-supported magnetized cylindrical disks.

This important study has inspired intensive research activity on the
linear MRI and its nonlinear development. This mechanism has
been investigated for various  physical conditions, geometries, and boundary
conditions (BC), and it has been approached using a number of
mathematical methods.  The complexity of its nonlinear
development requires employing numerical computational methods.
Understanding, or at least numerically simulating in a faithful way, the
{\em nonlinear}
transition and saturation processes (even through the use of some approximations),
is indispensable if the aim is to improve the more than 30 years old
phenomenological $\alpha$-model.

As the general problem of the development of magnetohydrodynmical (MHD) turbulence
in accretion disks is a formidable one, various approximations
have been utilized. The purpose of this paper is to assess the viability of
one of the
approximations most commonly used in numerical studies -
the {\em shearing box} (SB), also known as the ``shearing sheet."
The review articles by Balbus \& Hawley (1998) and Balbus (2003) contain
a comprehensive summary of the results of linear studies as well as
SB numerical ones (under various physical conditions) available at the time
of these reviews, together with a thorough list of references, of which only some
will be referred to here.

In the recent years, a number of new and improved (in resolution)
SB calculations subject to differing conditions and employing different numerical methods
have been reported on. Some results of these studies have raised a number of significant
issues regarding a central question pertaining to accretion disks: a quantitative estimate
and scaling of the angular transport in a saturated MHD turbulent state
(e.g, Pessah, Chan \& Psaltis 2007; Lesur \& Longaretti 2007, Fromang {\em et} al. 2007).
It is of no surprise that the estimate of this transport, or the "effective
$\alpha$", which is a quantitative measure thereof, must depend upon those relevant
nondimensional numbers characterizing the flow. In particular, these numerical
studies indicate that the transport is proportional
to some positive power of the magnetic Prandtl number of the flow.
\par

In parallel, a number of related approximate analytical treatments using asymptotic
analysis have been published (Knobloch \& Julien 2005; Umurhan, Menou \& Regev 2007;
Umurhan, Regev \& Menou 2007, hereafter URM07). The latter two studies found the
above-mentioned scalings in some limits of a magnetic Taylor-Couette thin gap
configuration (formally different from the SB only in the boundary conditions),
while the former work showed that transport diminishes in proportion
to the inverse of the Reynolds number and/or the magnetic Reynolds number
for the nonlinear development of coupled channel mode (CM) solutions.

Recent years have also seen experimental efforts to observe the MRI and its
nonlinear development in the laboratory.  These have been done
primarily in a magnetic Taylor-Couette (mTC) setup with an applied axial or
helical magnetic field. In these studies (see, e.g., Ji, Goodman \& Kageyama 2001;
Noguchi {\em et} al. 2002; Sisan {\em et} al. 2004; Hollerbach \& R\"udiger 2005;
Liu, Goodman \& Ji 2006; Stefani {\em et} al. 2006; and references therein)
MRI turbulence has only been reported for experiments with helical fields,
in which the azimuthal field component is significant.
Even for this case some controversy remains over the interpretation of the results
(Liu {\em et} al 2006; R\"udiger \& Hollerbach 2007).

Most of the existing SB calculations (at least of the early ones) employed a base state with
a constant vertical magnetic field and followed the development of perturbations
on this state. This situation, usually referred to as {\em non-zero}, or {\em fixed
net flux} (herafter fixed-flux), seems to be the most appropriate for disks threaded by
external magnetic fields (whether primordial or originating from the central accreting object).
The majority of existing numerical works of this sort have used {\em ideal}
(that is, non-dissipative) MHD SB equations with {\em periodic boundary conditions}
(periodic-BC) in the manner employed, for example, by Hawley et al. (1995),
but the very recent calculations of Lesur \& Longaretti (2007) include explicit
viscosity $\nu$ and resistivity $\eta$ in a small shearing box (SSB, see below),
high-resolution, numerical simulation.
Systematically changing the Reynolds number (${\rm Re}$) and the
magnetic Prandtl number ${\rm P}_m \equiv \nu/\eta$, in some range, these
authors aimed at
uncovering the trends and scalings of relevant physical properties, notably angular
momentum transport, with the relevant non-dimensional number(s).
A number of studies also exist demonstrating the role of the
MRI as an essential ingredient for internal dynamo action in the case where
the base state has {\em zero net flux} (hereafter zero-flux).
It is reasonable to suggest that the zero-flux case should be relevant
to disks (or disk regions) that are not threaded by magnetic flux.
The most recent zero-flux calculations in the SB framework with relatively
high resolution have
been reported  by Fromang \& Papaloizou (2007) and by Fromang {\em et} al. (2007)
These studies (both employing periodic-BC) focus on the issue of convergence
(i.e., numerical resolution) in ideal SB calculations and on the
$\rm Re$ and ${\rm P}_m$
dependencies of
the transport in dissipative conditions.

Recently King, Pringle \& Livio (2007), hereafter KPL07, pointed out
the discrepancy (of at least one order of magnitude) between the values of $\alpha$
inferred from observations and the estimates of this parameter based on
(mainly the early)
SB numerical simulations. In particular they reexamined the conclusion, reasoned by
Hawley, Gammie \& Balbus (1995) based on their fixed-flux SB simulations, that the effective
$\alpha$ is dependent (almost linearly) on the value of the imposed $B_z$.
Similar scaling has also been reported by Sano {\em et al.} 2004 and
Pessah, Chan \& Psaltis 2007, but the latter also found a pre-factor
depending on the box size as $L^{0.6}$.
They concluded
from this puzzling result that fixed-flux simulations are not likely to be adequate for accretion
disks and the zero-flux ones should be used instead. They also identified several theoretical difficulties
of the SB simulations - limitation of scale, problematic boundary conditions,
question of convergence of the simulations (numerical resolution issues), and a
possible breakdown of the MHD approximation.
Rincon, Ogilvie \& Proctor (2007) recently published a significant study, hereafter ROP07,
in which they discovered
that a self-sustaining nonlinear dynamo process may operate in Keplerian shear flows in
the zero-flux case. In this dynamo process, the
MRI is not a bona-fide linear instability, resulting from a modal stability analysis
of the base flow, but rather participates in one of the stages of the dynamo action.
We shall refer to the role of the MRI in this process as {\em mediating} rather
than ``driving", or ``inducing" (which will be the term in the fixed-flux case) MHD turbulence.
It is important to note, in the context of the present work, that this analysis
was done in the framework of a rotating magnetic plane Couette flow (rmpC), that is,
one in which realistic {\em wall boundary conditions} (wall-BC) were employed.
This is in contrast to the majority of SB calculations wherein periodic-BC are used
instead.

Turbulent magnetic dynamo theory, which is concerned with
the question of magnetic field generation in cosmic bodies
(see, e.g., Moffat 1978; Parker 1979), is a principally similar MHD problem and it
has been studied extensively
using direct numerical simulations (e.g, Cattaneo \& Hughes 1996;
Brandenburg \& Donner 1997; Schekochihin {\em et} al. 2007 and references therein).
The issues regarding the computational domain, resolution and  BC in such
calculations, and their effect on the results and findings
have always seriously been considered (see, e.g., Blackman \& Field 2000).
We mention this theory here at the outset because we will use it as an example of
another intensively studied MHD turbulence problem in which progress has been
quite impressive.

In this paper, we shall examine in detail the viability of
the SB approximation to the study of angular momentum transport (resulting
from MHD turbulence related to the MRI) in accretion disks. We shall
discuss it for both the fixed-flux case (where the MRI is expected to
induce MHD turbulence via a supercritical transition, i.e.,
it is a linear instability in the usual modal sense) and
the zero-flux case (where the MRI is invoked in a self-sustaining process
involving a subcritical transition).
We shall not only address and expound on some of the issues discussed by KPL07
and others,
but also point out and analyze in detail a number of other very significant
difficulties. This will lead us to the conclusion that the SB approximation, as used in
the great majority of the existing numerical simulations, suffers from limitations that
make it inadequate for the qualitative and quantitative understanding of the nonlinear
development of the MRI in accretion disks.

Our paper is organized as follows.
We start by reviewing a recent fairly exact
derivation of the SB approximation distinguishing between
the two limits of this approximation:
the {\em small shearing box} (SSB) and the {\em large shearing box} (LSB).
In the next section the relevant physical length scales of the problem
are discussed and related to the box scale, yielding
an answer to the question: which physical properties can
be expected to be faithfully uncovered in SSB and in LSB?
In Sect. \ref{resolution} we consider numerical resolution (convergence)
in SB simulations and its implications
on the numerical results and the conclusions thereof.
In Sect. \ref{BC}, we point out the
symmetry properties of the SB equations
and of the boundary conditions used in their simulations
(which are not globally valid in an accretion disk model)
and their effect on the solutions to the linear and nonlinear problems.
This is followed in Sect. \ref{energy} by a discussion of the SB
energetics and the effects of periodic boundary conditions on it,
when the box size is not large enough.
We also perform a number of numerical experiments
to demonstrate this important issue.

As far as investigating the properties of MRI induced and/or mediated
MHD turbulence is concerned, we finally {\em discuss} which (if any)
of the setups - SB, mTC or rmpC - would be an appropriate venue for
{\em local} numerical studies. In this context, we stress the
value of analytical and semi-analytical studies for
the purpose of better understanding numerical simulations and experiments.
We also suggest possible promising strategies for conducting
{\em global} (disk scale) studies, exploiting the knowledge
gained from the local ones.
The ultimate goal in all of this is to find a viable
way to quantitatively assess the turbulent
angular momentum transport and energy dissipation for
the modeling of accretion disks.

\section{The Shearing Box (SB) approximation}
\label{SB}

The essence of the SB approximation is in its {\em local} approach, that is,
the resulting equations for the perturbations on a steady base flow
are approximately valid in a small
region (a Cartesian box) around a typical point in the disk.
The {\em global} MHD base flow in an almost Keplerian accretion disk is not only
rotating and strongly sheared, but it is also inhomogeneous, non-isotropic
(endowed with non-zero gradients in density and other physical variables
and these gradients have very different scales in different directions) and
swirling (streamlines are curved).

The SB approximation, with its emphasis on the effect of shear,
is not new and has been introduced
in an astrophysical context long ago
by Goldreich \& Lynden Bell (1965) (see also Toomre 1964)
in their study of sheared gravitational instabilities in a galaxy.
Hawley \& Balbus (1991) adapted the approximation
to the numerical study of the MRI in accretion disks
and it has been routinely used ever since.
Umurhan \& Regev (2004) followed a systematic
derivation of the approximation, using asymptotic scaling arguments
with the purpose of quantifying the approximations that are
made leading to the SB (see Appendix A of that paper). Even though a purely
hydrodynamical case was considered there, the addition
of MHD terms is straightforward, because
the magnetic field remains invariant in a transformation to
a moving frame as long as the velocities involved are non-relativistic
(naturally with respect to an observer in the laboratory frame).

We summarize here the procedure for an MHD base flow in a thin
Keplerian disk with a constant vertical magnetic field (as a
convenient example).  This is done to
set the stage and also emphasize the difference between the two
self-consistent limits of the approximation.
The derivation starts by picking a point in the disk,
$(r_0, \phi_0, z=0)$ in cylindrical coordinates, transforming the
relevant full equation set into a frame rotating with the
Keplerian angular velocity appropriate for this point $\Omega_0
\equiv \Omega_{\rm K}(r_0)$, and looking at the equations in a
Cartesian box - a small neighborhood around the point $r_0$.
Non-dimensionalization of the equations that describe a thin disk
brings out two small non-dimensional parameters - $\epsilon$ (the
typical disk height $h_0$, in units of $r_0$) and $\delta$ (the
box size in the same units). The next step is the expansion (up to
first order in $\delta$) of the dependent variables of the laminar
(mean) base flow, Keplerian rotation in this case, in the
neighborhood of $r_0$, and the removal of the mean flow. This
results in a set of partial differential equations for the
disturbances that are fluctuations atop a base state (i.e,
deviations from a laminar flow profile with constant vertical
magnetic field). Identifying the radial (shear-wise) direction
with the Cartesian coordinate $x$, the azimuthal (stream-wise)
direction with $y$, leaving $z$ as the vertical coordinate and
neglecting the curvature terms (consistently with the smallness of
$\delta$) finally yields the SB equations. For details see Umurhan
\& Regev (2004), who identified two limits of the SB
approximation, depending on the ordering of the small parameters
$\epsilon$ and $\delta$. In the first one, a box whose size is
much smaller than the vertical scale height of the disk, i.e.,
$\delta\ll\epsilon$ is considered, and thus the unperturbed state
of linear shear may be considered homogeneous. The perturbations
are then incompressible and acoustics are filtered out. This is
the SSB. The second limit, LSB, is the case $\delta =
\order\epsilon$, in which vertical stratification as well as
compressibility must be taken into account and the situation is
considerably more complicated than SSB. We note here in passing
that a typical {\em thin} accretion disk has $\epsilon \sim
10^{-2}$, and this is thus the LSB ($\delta=\epsilon$) scale in
units of $r_0$, while the SSB ($\delta=\epsilon^2$) scale is in
these units $\sim 10^{-4}$.

For the purposes of this paper, it will suffice to consider here the
simple SSB for dissipative MHD flow with a fixed vertical background
field, possibly zero, and constant density ($=1$ in our units), because
the issues we want to deal with do not depend on additional
details. The velocity perturbations
(deviations from the laminar profile, in units of $\Omega_0 r_0 \delta$) are
${\bf u}\equiv(u,v,w)$ and those in the magnetic field (expressed in
units of $\tilde V_A^2$, a typical Alfv\'en speed squared, appropriate for
the background field $B_0$) are
correspondingly ${\bf b} \equiv (b_x, b_y, b_z)$. The total pressure
perturbation (in units of $\epsilon^2 \Omega_0^2 r_0^2$, see below)
is written as $\varpi \equiv p + (2\beta)^{-1} b^2$, where $\beta\equiv
\epsilon^2 \Omega_0^2 r_0^2/\tilde V_A^2$ (the
``plasma beta"). Additionally, lengths are scaled by the box size
($\delta r_0$) and time by the Keplerian time $(1/\Omega_0)$.
With these assumptions and definitions the non-dimensional
SSB equations in the rotating Cartesian frame explained above read,
\beq
\qquad \nabla\cdot {\bf u} = 0, \quad \nabla\cdot {\bf b} = 0,
\label{ssdiv}
\eeq
\beqa
(\partial_t - q \Omega_0 x \partial_y)u &-& 2 \Omega_0 v + {\bf u}\cdot\nabla u
= -\partial_x \varpi + \nonumber\\
&+& {1 \over \beta} \left( B_0\, \partial_z  + {\bf b}\cdot\nabla
\right) b_x + {1 \over {\rm Re}}\nabla^2 u
, \label{ssbx}
\eeqa
\beqa
(\partial_t - q \Omega_0 x \partial_y)v &+& (2 - q)\Omega_0 u + {\bf u}\cdot\nabla v =
-\partial_y \varpi +  \nonumber\\
&+&{1 \over \beta}\left( B_0\, \partial_z + {\bf b}\cdot\nabla \right) b_y
+{1 \over {\rm Re}}\nabla^2 v
,
\label{ssby}
\eeqa
\beqa
(\partial_t - q \Omega_0 x \partial_y)w &+& {\bf u}\cdot\nabla w =
-\partial_z \varpi +\nonumber\\
&+& {1 \over \beta}\left( B_0\, \partial_z + {\bf b}\cdot\nabla \right) b_z
+{1 \over {\rm Re}}\nabla^2 w
,
\label{ssbz}
\eeqa
\beqa
(\partial_t - q \Omega_0 x \partial_y) {\bf b} &=&
B_0 \partial_z {\bf u} + \nonumber\\
&+& q \Omega_0 b_x {\bf \hat y}+ \nabla\times({\bf u}\times {\bf b})
+ \frac{1}{ {\rm Re}_m} \nabla^2 {\bf b},
\label{ssinduction}
\eeqa
where the Reynolds number ({$\rm Re$}) and its magnetic counterpart,
${\rm Re}_m \equiv {\rm  Re \cdot P}_m$, are defined in the SSB
as:
\[
{\rm Re} \equiv \frac{\Omega_0 r_0^2\delta^2}{\nu},\qquad
{\rm Re}_m \equiv \frac{\Omega_0 r_0^2\delta^2}{\eta}.
\]
For the sake of some generality and convenience
we have kept $\Omega_0$ and $B_0$ (which are actually equal to 1 in our units) and
$q \equiv - (d\ln \Omega/d\ln r)_0$ (equal to $3/2$ for a Keplerian rotational law).
Note that to the extent to which these equations represent a small disk section,
if the base flow deviates
significantly from Keplerian rotation, then the horizontal pressure
gradient does not drop out of the equations and the system would not
be a proper representation of a rotationally-supported disk.
In addition, it is perhaps important to explain the choice of the pressure
unit. In a disk that is vertically supported by pressure, the vertical
scale-height is determined by the sound speed and gravity (that can be
expressed in terms of the local Keplerian speed). Thus at a point $r_0$, we obtain
that this vertical scale is approximately given by $h_0= \tilde v_s/\Omega_0$,
where $\tilde v_s$ is the typical sound speed. Therefore
$\epsilon = \tilde v_s/(\Omega_0 r_0)$ and it is small if the Keplerian
velocity is highly supersonic. As we have pointed out above, on the SSB
scale the flow is essentially incompressible and the sound speed loses
its meaning. By scaling the pressure with $\epsilon^2 \Omega_0^2 r_0^2$,
we thus consistently express the vertical size of the disk.

It is possible to reformulate the above SSB equations in
terms of {\it shearing coordinates} in the same way
as implemented by Goldreich \& Lynden-Bell (1965).
These coordinates are written into a frame that is shearing exactly
as the background flow itself on the SB scale (linearly).
This coordinate system $(\xi, \eta, \zeta, \tau)$ is thus defined
in terms of the coordinates in the rotating frame through the transformation
\beq
\xi = x, \quad \eta = y - q \Omega_0 t x,\quad \zeta = z,
\quad \tau = t,
\label{shearedmetrictransform}
\eeq
and then equations (\ref{ssdiv}-\ref{ssinduction}) may be
transformed in terms of the shearing coordinates
by only making the following formal replacements
\beq
\partial_t \rightarrow \partial_\tau  - q \Omega_0
\xi {\partial_\eta};~ \partial_x
\rightarrow \partial_\xi - q \Omega_0 \tau \partial_\eta;
~ \partial_y \rightarrow \partial_\eta
,~ \partial_z \rightarrow \partial_\zeta
\label{operatortransform1}
\eeq
and
\beq
 \nabla \to {\mathfrak{ O }} -
{\etaunit} q \Omega_0 \tau \partial_\eta,
\label{operatortransform2}
\eeq
where $ {\mathfrak{ O}} \equiv \xiunit \partial_\xi + \etaunit \partial_\eta +
\zetaunit \partial_\zeta $
is the gradient operator in the shearing coordinate system.

The LSB equations (both in terms of an observer
in the rotating frame and in terms of the shearing coordinates) are
somewhat more complicated
because the base state (background) vertical profiles
explicitly appear and the fluid must be treated as compressible.
Thus, equations of mass as well as
energy conservation have to be included, together with some kind of constitutive
thermodynamic closure relation (see Umurhan \& Regev 2004, for details).
If the LSB approximation is to be used in the context of an accretion disk
(this kind of approach is sometimes called {\em quasi-global}),
the vertical BC are no longer a matter of ``free" choice as they are actually
the physical conditions on the vertical edges of the disk.

We remark here, at the outset, that a significant
number of published SB numerical simulations seem to use neither
SSB nor LSB, but a kind of ``intermediate" approximation between
them. As the SSB and LSB are asymptotic representations of
thin-disks, it is not clear how much and which of the results
derived from these intermediate approximations are physically
relevant to disk systems. A number of examples of such seemingly
inconsistent calculations are given in the next section.

We finally note that virtually all
numerical calculations that utilize the SB approximation, in all
its variants, employ periodic-BC in which the periodicity in the
radial direction is sheared (see Hawley et al. 1995).  These
sheared-periodic-BC are equivalent to enforcing that all
perturbation quantities be triply (or in the case of a 2-D
calculation, doubly) periodic in the {\em sheared coordinates}
system. We shall discuss all these issues and their implications
in more detail below.

\section{Relevant length scales}
\label{scales}

In a typical accretion disk, $r_0 ( =1$ in units of $r_0$, which we shall use
in this discussion as the length unit) is the scale on which curvature
terms appear and this is also the scale of any underlying {\em radial}
structure gradients. Vertical stratification obviously appears on
the vertical pressure scale-height, which is $\epsilon = h_0/r_0 \ll 1$ in our notation
and units. This is also the scale on which the effects of compressibility are
manifested. These are the basic physical disk {\em structure} length scales.

In studies of turbulence it is customary to consider the {\em injection} scale,
$\ell \equiv k^{-1}_{\rm in}$, on which the process of energy injection into the
system is effected and the {\em dissipation} scale, $\equiv k^{-1}_{\rm d}$,
on which microscopic viscosity (and in MHD also resistivity)
are operative. These are linked by the {\em inertial} range of scales.
Since  $k_{\rm in}\ll k_{\rm d}$ (at least in astrophysical systems),
the inertial range extends over many orders of magnitude. In MHD turbulence,
the energy cascades in 2D and 3D
are both direct, i.e., from small wave-numbers to large ones (see, e.g., Biskamp 2003),
and consequently, the injection scale can be roughly identified with the system's structural
scale. In the classical Kolmogorov theory, the scale of the largest eddies,
which contain most of the energy, is referred to as the  {\em integral} scale.
This is also the length scale appearing in the definition of the Reyonlds number
and serves as as a measure for the scale over which the turbulent fluctuations
are correlated. To avoid complications we shall practically identify the integral scale
with the correlation length and the injection scale, denote them as $\ell$,
and refer to them interchangeably. This identification has an obvious intuitive
physical basis.
\par
In a disk that is strongly non-isotropic, two disparate
injection scales seem a priori to appear - the horizontal one
$\ell_{\rm h} \la  r_0$ and the vertical one $\ell_{\rm v}
\la \epsilon r_0$ (both given here in dimensional units). It is, however,
reasonable to identify $\ell$ with the smaller of the two, because
this is the size of largest eddies. A quantitative measure of
$\ell$ can be "phenomenologically" determined by the relevant
wave-number at which the disturbance kinetic energy $E_{\rm k}(k)$
spectrum peaks. The fact that the scale at which resistivity and
viscosity come into play need not be equal (${\rm P}_m\equiv {\rm
Re}_m/{\rm Re}$ is not generally of order 1) also makes the
determination of the dissipation scale not unique.
In accretion disks, typically ${\rm P}_m < 1$,
i.e., $k_\nu > k_\eta$ where $k_\nu$ and $k_\eta$, respectively,
represent the dissipation scale wave-numbers for viscosity and
resistivity.  In thin accretion disks, especially if the
ionization is not complete, these inequalities are strong.
\footnote{Note that in cold disks it is expected that ${\rm P}_m
\ll 1$ throughout the bulk of the structure.} Phenomenologically,
the inertial range is considered to be the $k$ region, in which
$E_{\rm k}(k)$ exhibits a power-law behavior. Thus, the inertial
range in the MHD turbulence of a disk can be considered as
including wave-numbers roughly in the range $\epsilon^{-1}\ll k
\ll k_\eta$. This may be further complicated by the fact that MHD
turbulence can be {\em inherently} non-isotropic, even if
the system is structurally isotropic (the mean magnetic field
breaks the symmetry) and one should consider $k_\|$ and $k_\perp$
separately (this is the case, for example,
in the Goldreich \& Sridhar theory of interstellar MHD turbulence)
For a fairly detailed discussion of the  MHD turbulence length scales
and relevant references see Biskamp (2003).

If a local (SB) calculation is performed, the scale of the system is the box
scale, which is $\delta\ll\epsilon$ for SSB and $\delta \sim \epsilon$ for LSB.
In this context, even though it is evident that the equations resulting from the SB
approximation possess a radial (i.e. in the shear-wise direction $x$)
scale invariance, computations done on the equations resulting from the SB approximation
clearly do introduce a length scale, which is the size of the box.
Thus, the spatial scale invariance symmetry of the governing equations is
lost in the solutions whenever a computation is performed. Moreover,
any analysis done within the SB approximation
can be meaningful only if the length scale of the physically relevant structure
(e.g. the most unstable linear mode, the injection scale) is significantly smaller than
the box size (which we shall refer to as $L$, in dimensional units).
In particular, {\em nonlinear} analysis (e.g., performed by numerical simulation) will not be
able to capture possible nonlinear coupling between the
unstable linear modes with modes whose
wavelength scale is larger than $L$ (see also the discussion at the end
of Sect. \ref{BC}). It appears that Goldreich \& Lynden-Bell (1965) were well aware
of this fact and indeed in their nonlinear analysis they returned to the full equations
of motion in rotating coordinates.

Local linear stability analysis (see Balbus \& Hawley 1998) performed
on a thin ($h_0/r_0 = \epsilon \ll 1$) Keplerian disk
with disturbances assumed to have the form $\propto \exp(i{\bf k\cdot x})$
(i.e., actually having no structure)
predicts for the maximally unstable MRI mode ${\bf k}_m\cdot {\bf V}_A=\Omega_0 {\sqrt{15}/4} $,
with ${\bf V}_A \equiv {\bf B}_0/\sqrt{4 \pi \rho}$ being the Alfven velocity
defined here in dimensional units.
For a vertical background field we get for the vertical wavelength of the most unstable mode,
$\lambda_m$,
\beq
\frac { 2\pi}{k_m}=\frac{8 \pi} {\sqrt{15}}~ \frac{V_A}{\Omega_0}=
\frac{8 \pi} {\sqrt{15}}~\frac{V_A}{\tilde v_s}~ h_0 = \epsilon \frac{8 \pi}{ \sqrt{15 \beta}}~
 r_0 = \frac {\epsilon}{\delta}~\frac{8 \pi}{ \sqrt{15 \beta}}~L.
\eeq
Here, the thin disk relation $h_0 = \tilde v_s / \Omega_0$ and the
definition of the box size $L=\delta\cdot r_0$ have been used.

In order to capture the fastest growing mode in the SSB ($\delta \ll \epsilon$)
the plasma beta has to be very large, explicitly $\beta \ga 40 \epsilon^2/\delta^2$.
Sure enough, modes with vertical scale smaller than $\lambda_m$ are also unstable
and therefore any numerical calculation within the SSB scheme with $L<\lambda_m$
and with periodic-BC (see below) should
display the mode whose wavelength is equal to the box size as the most unstable one.
Indeed, typical SB simulations (see Hawley \& Balbus 1989) clearly
display this feature (as a transient),
with the CM (also referred to as "channel flows", see also below)
breaking up later on, {\em nonlinearly}. If the
SB size, or alternatively $\beta$, is not large enough, so as to allow for the capture
of the entire unstable part of the spectrum, it cannot be a priori expected that
the simulation yields reliable results.
In the LSB scheme (with the size of the box of the same order of magnitude as the disk
thickness), the most unstable linear mode will be contained in the box
for reasonable values of $\beta$. But, a calculation of this sort is more involved -
vertical structure as well as compressibility have to be included.
The inclusion of dissipative effects, i.e. finite ${\rm Re}$, ${\rm Re}_m$,
further changes the structure of the dispersion relation
and a judicious choice of these parameters may ``push"
the unstable part of the spectrum down to within the box size,
but the large majority of SB calculations for the fixed-flux case
have been with ideal MHD.

In any case, possible nonlinear interactions between the short vertical wavelength unstable
modes, with modes (possibly also horizontal ones) whose scale is larger than the SB scale,
cannot be captured in a SB simulation. It is conceivable that such interactions
may play a role in the instability saturation and the development of activity.
In the turbulent dynamo problem
such an interaction between scales is seriously considered (e.g., Cattaneo \& Tobias 2005).
We shall also see in Sect. \ref{BC} that when one examines a perturbation
consisting of a wave-packet (rather than just one mode) the situation becomes
quite complex, even in linear analysis.

Moving now to a discussion of the small (dissipative) length scales,
we note that the resistive and viscous scales in accretion disks are
extremely small in comparison to the box scale, even in the SSB case. A numerical
SB calculation faithfully resolves scales between the system's (box) size and
some lower limit $l_{\rm num}$, the latter of which results from
the finite accuracy of the numerical
scheme. It is easy to see that
in all existing SB simulations $l_{\rm num}$ is much larger than any
realistic dissipation scale (for an accretion disk). In ideal MHD
calculations, the scale $l_{\rm num}$ clearly plays
the role of the dissipation scale, but it is quite doubtful that such numerical
dissipation faithfully reproduces the correct physical dissipation, and so
it may conceivably influence the results on larger scales.
This issue will be discussed
in more detail in Sect. \ref{resolution}.

We conclude this discussion of relevant scales by pointing out
again, and giving some examples of, the above mentioned
apparent inconsistency in some SB numerical simulations that use
neither SSB nor
LSB, but a kind of ``intermediate" approximation between them.
Balbus, Hawley \& Stone (1996) include compressibility
but no vertical structure (LSB has to include both, and SSB none),
while Sano \& Stone (2002) include compressibility and investigate
further physical ingredients (the Hall effect), but also take the
background structure as vertically uniform. Similarly, the
simulations of Papaloizou \& Fromang (2007) and Fromang {\em et}
al. (2007) use
the LSB (box size equal to the disk thickness), but the vertical
background is uniform. In all these calculations periodic-BC are
employed in the vertical direction. Moreover, the initial
conditions of the perturbations in the last two calculations
involve a sinusoidal vertical field in the shear-wise direction
$x$ with an amplitude that makes the magnetic energy density in
the perturbation not infinitesimal with respect to the pressure.
It means that this is a case where finite-size (nonlinear)
perturbations are causing a subcritical transition to turbulence.
This is because under these viscous and resistive conditions there
is no exponentially growing normal mode instability but there is,
instead, strong transient growth (cf. ROP07). It is not clear how
simulations considering a medium whose vertical
structure deviates substantially from that of a disk,
have insufficient resolution in the ideal case (as shown
by the authors themselves), and employ periodic-BC in the
vertical, can faithfully uncover the extremely intricate processes
that must be operating in such transitions (cf. ROP07).
We think that it would be fair to say that,
though the results are interesting and worth studying
in their own right (at least from the standpoint of
understanding the interplay of the mechanisms at work), the applicability and relevance of
these results to the
physical conditions characterizing real disks
is doubtful. In contrast, the
simulations of Lesur \& Longaretti (2007) are done in the SSB
model (incompressible flow, constant background) and periodic-BC
in all three directions. This is consistent according to our view
of SSB vs. LSB expressed above, but the periodic-BC, when applied
in a SB of too limited a size, may
still pose a difficulty (see below in Sect. \ref{BC}), and it
would thus be advisable to experiment with different boundary
conditions as well.

It is not yet completely clear how the inconsistencies and assumptions
discussed above, in the models used thus far,
may influence the results and how they will ultimately
relate to real thin accretion disks (or at least large sections of them).
However, given the concerns we raised thus far
we think that these matters must be openly reconsidered.
Certainly a lot more comparative numerical experiments
are necessary before a better understanding can be achieved.
It is our opinion that if the goal is to analyze the
effects of compressibility and vertical structure
in a sectional representation of a real thin accretion disk
then the LSB should be used together with a physically meaningful
prescription for the boundary conditions in the vertical direction.
The LSB by its very derivation includes these effects in a
physically and mathematically consistent way. In contrast, if
the SSB (having a uniform background density and pressure) is used,
there is no need  to simulate a compressible flow, which is computationally
more costly and physically more involved (e.g., it has to be monitored
to insure the smallness of the Mach number), even though it is not incorrect, per se.

\section{Numerical resolution and its relationship to dissipation and saturation}
\label{resolution}

Astrophysical fluid systems and their magneto-fluid counterparts (like an accretion disk flow)
are endowed with extremely large Reynolds ($\rm Re$) and magnetic Reynolds
(${\rm Re}_m$) numbers and therefore
numerical simulations cannot yet resolve the full dynamical range of
such turbulent flows.
In other words, given the present magnitude
of computing power,
numerical resolution of the full spatial range of these systems, from the system's
energy injection scale through the full inertial scale and down to the dissipation
scale, is still out of reach.

It has been argued in the past (e.g., Balbus \& Hawley 1998) that
perhaps a full dynamic range is actually not needed in the accretion disk
problem because the scale of the most unstable {\em linear} MRI modes
is large and, in the saturated turbulent state, most of the energy resides
in, and the angular momentum transport is done by, the large scale eddies.
Moreover, it has been explicitly stated in that paper (basing the argument
on the numerical work on hydrodynamic turbulence of Oran \& Boris 1993)
that the physical nature of
the turbulence on large scales will not change if one lets
``numerical dissipation" cut off the eddy energy spectrum on the grid-scale
$l_{\rm num}$, instead of allowing for the full inertial range, down to the
true physical dissipation scale.

This conjecture is debatable for hydrodynamical flows
(see, e.g., Pouquet, Rosenberg \& Clyne 2002), even for $l_{\rm num}$
only modestly larger that the dissipation scale and certainly for conditions
appropriate for accretion disk flows. For an extensive, up-to-date,
discussion on the use of unresolved hydrodynamical codes for the study of
turbulence, see Grinstein, Margolin \& Rider (2007).
In any case, the conjecture appears to be
clearly unfounded for MHD turbulence with very large Reynolds numbers.
Boldyrev \& Cattaneo (2004) found
that numerical or experimental investigations of MHD turbulence
(at least for small ${\rm P}_m$, as is the case in most disks)
need to achieve extremely high resolution to describe magnetic
phenomena adequately and Schekochihin {\em et} al. (2004, 2005, 2007)
have addressed in detail the question of the relevant scales for the various
cases (values of ${\rm Re}$ and ${\rm P}_m$) and the resulting
computational challenge (the need for extremely high resolution).
These works deal with the turbulent dynamo problem, which
is in principle related to the accretion disk problem
(at least in the zero-flux case).
Similar general conclusions also appear in Biskamp (2003).

Thus, it appears that a viable estimate of $\alpha$ in a
MHD turbulent disk cannot be obtained from poorly resolved SB simulations.
In addition, it is even difficult to quantitatively assess the energetics
of MRI driven turbulence from such simulations. Gardiner \& Stone (2005)
investigated this issue in detail (see also Sano \& Inutsuka, 2001)
using an ideal compressible SB calculation with fixed-flux. The dissipation
and hence the thermal energy was calculated by relying on ``numerical" viscosity
and resistivity (Sano \& Intsuka 2001 used an explicit prescribed resistivity).
If the nature of the turbulence on scales resolved by these calculations
(i.e, larger than $l_{\rm num}$)
{\em does} indeed depend on the unresolved small scales (see above),
the energetics and transport properties should be affected as well.
The numerical investigations of Pessah, Chan \& Psaltis (2007) and of
Fromang \& Papaloizou (2007) actually show, for the fixed-flux case
that increasing the resolution of simulations relying on numerical dissipation
shows a decreasing trend in the calculated transport, suggesting
that this is indeed the case.

The saturation of the linear MRI (a
supercritical transition to turbulence in the fixed-flux case),
must be effected by dissipation if the resulting state of the SB
is to represent a thin Keplerian disk. Saturation, in general, can
happen either by the ``removal" of the cause of the linear
instability and/or by the dissipation of the energy driving the
linear instability.  For the MRI ``removal" would be either the
expulsion (or redistribution) of the background magnetic flux,
or the neutralization
of the destabilizing shear profile.  Beginning with the latter we
note that the gravitational field of the central object, which
causes the Keplerian shear in the disk, can obviously not be
annihilated by any internal processes in the disk. Neutralization
of the shear profile can only occur if a new
azimuthal velocity field emerges whose profile masks the
destabilizing effect of the background shear. The aggregate flow
field could then be such that the fluid is then marginally stable
to the MRI. However, this would imply a significant deviation from
a Keplerian profile and the resulting state of the system would no
longer represent a rotationally supported thin disk because there
would now emerge significant radial gradients of the pressure.
With regards to the former saturation option we note that the
surface integrated magnetic flux in the SB remains a constant.
Therefore, setups considering a constant-vertical field cannot
vertically expel that uniform field (processes resulting from
buoyant instabilities of the horizontal fields inside the disk are
still possible, as well as possible redistribution
of flux). Thus, a fully turbulent state of MRI-driven
turbulence must achieve saturation by disposing of the energy
gained during the linear instability stage by dissipation at the
smallest scales, if it is meant to represent the saturated
turbulent state of a thin disk. It is thus difficult to see how an
ideal MHD simulation that does not resolve the dissipative scales
can faithfully represent the process of turbulent dissipation and
angular momentum transport in such a MHD-turbulent state of a
fixed-flux disk. The work of Fromang \& Papaloizou (2007) suggests
that this is the case in a zero-flux system as well.
Thus proper treatment of the microscopic dissipation seems to be imperative.

Numerical modeling of turbulent flows
is obviously not exclusive to astrophysics and a large body of work and
literature exists on this subject (e.g., Pope 2000, Davidson 2004,
and references therein).
Among the basic
approaches are {\em direct numerical simulation} (DNS) and {\em large
eddy simulation} (LES), where in the former approach all scales (up to the numerical
domain size, of course) are calculated (and well resolved)
employing some finite Reynolds numbers, while in the latter approaches
scales below $l_{\rm num}$ are treated with the help of a turbulence model.
The books by Wilcox (1998), Dubois, Jaubereau \& Temam (1999), Pope (2000) and
Davidson (2004) discuss
in detail these approaches and their subtleties as well as other turbulence models.
Problems encountered in engineering applications, as well as those addressed by the
fluid-dynamical community in general, opt for the appropriate numerical scheme,
according to the problem at hand. The guiding principle in making the
choice of the method is usually the desire to introduce the minimum amount of complexity
while capturing the essence of the relevant physics.
In astrophysical flows in general and in the accretion disk problem in particular,
the choice of the BC introduces an additional difficulty for faithful numerical
modeling.

We find it useful to classify
the following three different numerical approaches to the
problem of nonlinear development of MHD turbulence in
accretion disks.
\begin{enumerate}
\item
Local ideal and dissipative calculations - with the purpose of
extracting the maximum information on the local properties of the
turbulence in magnetized rotating shear flows and obtaining
some quantitative
measure of the turbulent transport coefficients (in particular).
These are a DNS-like, maximally resolved, calculations.
\item
Global calculations employing some physically motivated sub-grid scale turbulence
model. Such models can fully include mass transfer as well as other
important properties of accretion disks. In principle, these allow confronting
accretion disk models with relevant observations. These are a LES-like
calculations.
\item
Global calculations, albeit with unrealistically low Reynolds numbers, -
in the purpose of discovering the trends of the dynamics with increasing
${\rm Re}$ and/or ${\rm Re}_m$. In a way, this approach is intermediate
between the former two.
\end{enumerate}

All existing SB simulations nominally belong to the first class,
however the great disparity between the smallest resolved scale
$l_{\rm num}$ and the true dissipation scales prevent them from being
regarded as true DNS. There exist also other studies that address
the nonlinear development of the MRI by utilizing
rotating magnetic plane Couette (rmpC),  whose equations
are formally equivalent to the SB equations, or
magnetic Taylor-Couette (mTC) setups. These are flows confined
between ``walls" on which specific well-defined boundary
conditions (wall-BC) are employed and they have been
mainly aimed at laboratory experiments of the MRI. In some
of these calculations spectral schemes with very high resolution
have been used (e.g., Obabko, Cattaneo \& Fischer 2006) so as to
achieve a true DNS status for relatively very large values of
the Reynolds numbers.

At present, existing global or quasi-global (LSB) calculations
all suffer from insufficient resolution to achieve the status of a DNS simulation
of accretion disks with ``astronomical" Reynolds numbers.
Except for the old $\alpha$ prescription for MRI mediated or induced  MHD
turbulence, there is as yet
no thoroughly tested sub-grid model as might be employed in a LES
(or one using some other turbulence model) calculation.
The MHD turbulence model developed by Ogilvie (2003) and the model for
hydrodynamic turbulence in Couette flow (as in the recent work by Dubrulle et al. 2005
on strongly rotating flows) appear promising.  Thus far, such models
(or any other for that matter) have not
been employed
in simulations pertaining to accretion disks. Moreover,
turbulence models are usually tailored to a particular problem, exploiting
relevant results of laboratory and/or numerical experiments. The former
model used the results of SB simulations, which we discuss in this paper
and doubt their reliability, while the latter model is probably inappropriate
for MHD flows.

Among such global calculations Gammie \& Balbus (1994)
investigated the linear stability in an ideal LSB (with
periodic-BC in the shear-wise direction and various choices for
the vertical BC) and found that the BC play an important role.
Stone {\em et} al. (1996) performed a numerical simulation for
these conditions, obtaining a magnetically dominated disk with a
corona and estimated the value of the effective $\alpha$
parameter. However, it is difficult to confidently
evaluate this result in light of our previous comments - in
particular because the numerical resolution seems insufficient and
periodic-BC are used in the vertical direction. Hawley (2001)
performed a global numerical calculation on a cylindrical domain,
but used a particular artificial viscosity prescription and it is
difficult to say much about the effective Reynolds numbers in this
calculation (except that they are low and uncontrolled).

The work of
Kersal\'e {\em et} al. (2004), who investigate a global linear problem for a model
cylindrical incompressible flow testing the effects of a variety of BC,
including those that bring about unstable wall-modes, can be categorized
as belonging to the above third approach.  In the follow-up
work Kersal\'e {\em et} al. (2006) use a spectral code to
examine the nonlinear development of the wall-modes
by adopting a well defined (if, however, relatively low) ${\rm Re}$ at ${\rm P}_m=1$.
This approach has at least the advantage that dissipation is controlled
and there may be hope that both the linear and nonlinear trends
of important flow properties can perhaps be asymptotically uncovered for
realistic Reynolds numbers (like, e.g., in the local
hydrodynamic study of Lesur \& Longaretti, 2005).

Given the doubt discussed above that numerical dissipation by itself
is sufficient in an LES calculation, it seems to be imperative to
look for a sub-grid model that is appropriate for the problem at hand.
Investigating MRI induced or mediated turbulence in local DNS simulations
may certainly offer in-roads into a better understanding of
its {\em local} salient physical features, which include, e.g.,
some characterization of the emerged turbulence in terms of eddy spectra and correlations.
Information of this sort can provide the ingredients for the needed ``turbulence
model," which would go into a faithful global accretion disk LES
(or a simulation employing other turbulent flow computational schemes).
Since local simulations are needed to infer the turbulent behavior down to the dissipation
scale, rmpC with physically definite boundary conditions seem to be not less appropriate
than SB (whose periodic-BC introduce additional difficulties
and inconsistencies, see below) for this purpose.
Of course, such simulations
must be fully resolved (DNS) and pushed to the maximum $\rm Re$ and ${\rm Re}_m$
values in order for them to yield this kind of information.

The present limitations on computing power and architecture
currently rule-out making progress through global DNS simulations
of accretion disks. Global simulations with unrealistically small
Reynolds numbers are a possibility, but it is still unlikely that
one can learn much about the turbulent angular momentum transport
occurring under real accretion disk conditions.  However, it is a
step closer in that direction and it seems (to us) to be a
worthwhile tract to follow this strategy.

\section{Symmetry and boundary conditions}
\label{BC}
The SB approximation with periodic-BC is endowed with a particular important symmetry:
it allows for shear-wise (in the Cartesian geometry of the box it corresponds
to the $x$-direction) invariant solutions.
The CM are solutions of the linear problem in the fixed-flux case
and they are unstable (exponentially growing) for wave-numbers $k_z$ below some critical
value for a given $B_z=B_0$. Moreover, these linear modes individually are also exact
solutions to the nonlinear SB equations, that is, they can appear and exponentially grow
(on the rotation time-scale) beyond any bound. Goodman \& Xu (1994) showed
that these growing modes become linearly unstable
and eventually break up.
This work has served as being an important step toward {\em understanding}
the transition from linear MRI to turbulence in accretion
disks in the fixed-flux case
(see Balbus \& Hawley 1998).

The $x$-invariance symmetry, which allows for the existence of CM,
is obviously broken by any non-periodic-BC on $x$
even in local SB analyses and this is manifested on the box scale.
As discussed in Sect. \ref {scales}, curvature terms in swirling flows
(like accretion disks or mTC) as well as radial structure gradients,
which also break the above symmetry, are global.
The SB scale, appropriate for thin accretion disks satisfies
$\delta \cdot r_0 <L < \epsilon \cdot r_0$, while
radial symmetry breaking by the above mentioned global
physical effects are manifested on a length scale of the order $r_0$.
This symmetry breaking can thus be ``communicated" through a real
compressible disk flow only on timescales larger than the horizontal sound-crossing time,
$\tau_{\rm sound} \sim r_0/v_s \sim (r_0/h_0)/ \Omega_0$.  On the other
hand the (linear) growth time  of the CM is
$\tau_{\rm CM} \sim 1/\Omega_0$. Thus, because $r_0/h_0=\epsilon^{-1}
\sim 50-100$ in most thin accretion disks, one might think that the
broken symmetry can not affect the development of the CM. However,
the saturated MHD turbulent state in the disk, postulated to result from the
Goodman \& Xu (1994) CM instability, must persist for very long times
(relative to even $\tau_{\rm sound}$) and so this global state
has to be driven by a continuous nonlinear process, giving rise
to energy injection along the full global scale of the disk.

It is important to mention, in this context, the work of
Coppi \& Keyes (2003), who showed that in a disk of finite vertical
extent, in which the modes have to localized vertically (and not an
infinite cylinder or a finite one, but with periodic BC), the modes
necessary acquire a characteristic oscillatory profiles in the radial
direction, i.e., are certainly not of the CM type and have much longer
rise-times.
This is probably related to the very
recent investigation by Liverts \& Mond (2007), who examined whether in
the simplest ideal MHD setting, with the only spatial dependence
of the perturbations being $\propto e^{i k_z z}$ (like of the CM),
an {\em absolute} or perhaps {\em convective} linear instability
(see, e.g., Schmid \& Henningson 2000, \S{7.2}, for a detailed
discussion of this topic) occurs. Solving analytically an initial
value (Cauchy) problem for the MRI in the fixed-flux case, with an
initially localized perturbation in the form of a Gaussian with
prescribed width, they found that the
spatio-temporal evolution of the packet is not as trivial as that
of single CM or a superposition thereof. The initial perturbation
was found to split into two wave packets moving in opposite
directions (vertically) from the initial position. As each
wave-packet propagates at its group velocity (of the order of
the Alfv\'en speed), the amplitude of the packets does not grow in
time. Instead, it oscillates sinusoidally in time in the co-moving
frame of the wave-packet.  In addition to the two
oppositely moving wave-packets, Liverts \& Mond found an
absolutely unstable singular component, resulting from the
well-known branch point of the dispersion relation
$(\omega_b,k_b)=(3i/4, \sqrt{15}/4)$, growing in time
asymptotically (that is for large enough $t$)
like $\propto e^{\gamma t}/\sqrt{t}$, where $\gamma \equiv |\omega_b|$
and having an amplitude that is a function of the initial packet width.
Consequently, the linear stability analysis yields a situation
considerably more complicated than just of exponential growth of
CMs, with their subsequent secondary instability of Goodman \&
Xu (1994). In particular, since the disk is not infinite in its
vertical extent, reflections of the wave packets by the boundaries
may complicate the situation. The absolutely unstable component
does not grow fast enough before the reflected packets return to
their origin and therefore periodic-BC are simply inapplicable.
Thus, the quite simple and elegant
Goodman \& Xu {\em local} linear scenario, which is generally accepted as
an explanation of the mechanism for transition from linear MRI to turbulence
in a fixed-flux disk, should be reconsidered.

Linear analysis of a {\em global} configuration does not
show the presence of CM (see the discussion
of ``body modes" in Kersal\'e {\em et} al. 2004).
Moreover, the nonlinear development also appears to be quite different
(see Kersal\'e {\em et} al. 2006).
Only a global calculation, of the kind
performed by Kersal\'e {\em et} al. (2006), for example,
but allowing for a {\em compressible} flow and definite
{\em vertical} BC, can perhaps clarify this point.
In any case, it seems that the role of the CM in transition to
global MHD turbulence in an accretion disk threaded by a vertical
magnetic field is quite academic. As important as such {\em local}
understanding may be, it is not clear if it has any relevance to the
global problem.

Turning now to a discussion of the effects of periodic-BC, we first note
again that the majority of numerical studies on the nonlinear
development of MHD turbulence in accretion disks, aiming at the understanding
of the properties of the saturation and transport, employed some version of
the SB approximation. The periodic-BC employed in virtually all SB calculations,
have several adverse side-effects.
We shall distinguish here between the periodic-BC on $x$ in sheared coordinates
and the vertical (on $z$) periodic-BC.

It was already mentioned in the previous section.
that the $z$ periodic-BC may cause the inability to capture the most unstable mode.
But this can be remedied by either making $\beta$ very
large (i.e. a very small magnetic field) or by including dissipation and choosing an appropriate
combination of ${\rm Re},{\rm P}_m$, and $\beta$ \footnote{Or, alternatively, the
the box Cowling number, ${\rm Co} \equiv \tilde V_A^2/\Omega_0^2 L^2 =1/\beta$}
(see Umurhan, Menou \& Regev 2007). However, $z$ periodic-BC  for the fixed-flux case
are bound to cause un-physical results, for at least two reasons. First, they would give rise
to a spurious reentry of any convected component of the perturbation, when it reaches
a boundary (Liverts \& Mond 2007, see above). Second, magnetic fields, as is well-known,
have a necessarily {\em global} nature. This means that {\em any} local
calculation, in which well-ordered magnetic field lines extend beyond the computational
domain, must employ BC based on the knowledge of the magnetic configuration outside
the domain. Thus, arbitrary vertical BC (in particular $z$ periodic-BC) on the field (in
the fixed-flux case) will necessarily exert extraneous magnetic stresses on the system and may
substantially alter its dynamics. This point has recently been brought to attention
by Shu {\em et} al. (2007).

We finally turn to the most important observation about the use
of periodic-BC in numerical studies of turbulence and its bearing
on SB simulations.
Clearly, space periodic-BC are not accessible in
realistic physical situations, and they only can be faithfully employed
in the study of homogeneous turbulence when one assumes that real
boundary effects are not important (Dubois, Jauberteau \& Temam, 1999).
The caveats pertaining to the use of periodic-BC in numerical simulations
of turbulence are discussed in detail in \S7.2 of Davidson (2004) from
which we only bring here
issues that are explicitly relevant to SB simulations of MHD turbulence in an
accretion disk.
In order for the bulk of the turbulence not to be seriously affected by the
imposed periodicity, the box size should satisfy $L \gg \ell$, otherwise the
"tails" of the relevant correlation functions (of the turbulent fluctuations)
are not only cut off, but also very significantly lifted (see figure 7.6 of
Davidson 2004). This crucially affects the behavior of the {\em large} scale
dynamics. Thus, SSB is clearly out of the question if the aim is to gain
any understanding of accretion disk turbulence up to the injection scale.
 If we accept the reasonable
conjecture that the injection scale relevant for an accretion disk is indeed
$\ell=\ell_{\rm v} \la h_0$, the {\em minimal} size of the SB must be the
disk height. But then $z$ periodic-BC cannot be used and we are only left with
the option of semi-global (LSB) or global simulations. SSB simulations can
still be useful, but only for understanding the dynamics on very small scales, with
the purpose of devising an appropriate sub-grid model for global, or at least
semi-global LES. In the next section we shall demonstrate, with the help of a
numerical experiments, that using a SB, whose size is not large enough (as compared
to the scale of the relevant dynamical structures), and periodic-BC, gives
rise to spurious energy fluctuations, casting a serious doubt on the physical
significance of the simulation results.

\section{Energetics of the SB}
\label{energy}
\subsection{Integral statements}

In any continuum mechanics problem posed in a finite domain, as is the case
for SB MHD numerical calculations of the kind discussed here, it is generally
advantageous to consider integral physical conservation statements, valid on
the domain. A particularly useful consideration of this kind, in this problem,
arises from the examination of the energy budget for the system of equations
(\ref{ssdiv}-\ref{ssinduction}).
Using the definitions ${\bf B} \equiv \{b_x,b_y,B_0+b_z\}$, the total magnetic field vector
and ${\bf U} \equiv \{u, U_0+v, w\}$, the total velocity vector, where the
background field $B_0$ may be equal to zero (for the zero-flux case) and the
the background shear is embodied in $U_0 \equiv -q\Omega_0 x$, one may take the appropriate
inner products with the equations of motion and integrate them over the SB.

After rather standard manipulations the following {\em total} energy theorem results
\beq
\frac{d}{dt}\mathfrak{E}
= {\cal S} - {\cal D},
\label{total_E_int}
\eeq
in which
\beq
\mathfrak{E}\equiv \int {\cal E} d^3{\bf x}
\eeq
is the integral over the SB of the total energy density $\cal E({\bf x, t})$, defined
to be
\beq
{\cal E} =
\frac{1}{2}{\bf U}^2 +
\frac{1}{2\beta}{\bf B}^2
-q\Omega_0^2 x^2 =
E_{u}+E_{{\rm shear}}+ E_B - q\Omega_0^2 x^2,
\label{total_E_def}
\eeq
where $E_{u} \equiv {1 \over 2} |{\bf u}|^2$ is the kinetic energy density of the {\em perturbation}
atop the linear shear $U_0$, $E_B \equiv \frac{1}{2\beta}{\bf B}^2$ the {\em total}
magnetic energy density, $E_{{\rm shear}} \equiv -q\Omega_0 x v + \frac{1}{2}q^2\Omega_0^2 x^2$
and we have explicitly used $U_0=-q\Omega_0 x$.
The term $-q\Omega_0^2 x^2$ in the definition of ${\cal E}$
is the centrifugal potential energy density, arising in a SB rotating with the local
Keplerian velocity
(referred to as ``tidal"-potential by
Gardiner \& Stone, 2005). It obviously merely represents a local Taylor expansion of the
gravitational field potential upon which the fundamental shear in this flow
derives its existence.

The terms on the righthand-side of equation (\ref{total_E_int}) are
\beqa
{\cal S} &\equiv&
-\oint {\hat {\bf n}}\cdot {\bf U}({\cal E}+\varpi) d{\bf \sigma}
+
\frac{1}{\beta}\oint {\hat {\bf n}}\cdot {\bf B}({\bf B}\cdot {\bf U})d{\bf \sigma}
+\nonumber\\
&+&\frac{1}{2{\rm Re}}\oint ({\hat {\bf n}}\cdot \nabla) {\bf U}^2d{\bf \sigma}
+\frac{1}{2{\rm Re}_m}\oint ({\hat {\bf n}}\cdot \nabla) {\bf B}^2d{\bf \sigma},
\label{S_calE_definition}
\eeqa
where ${\hat {\bf n}}$ is the unit normal of the bounding surface of the box and $d\sigma$
is the differential area element of the bounding surface,
and
\beq
{\cal D}
\equiv {\frac{1}{{\rm Re}}}\int |\nabla {\bf U}|^2 d^3x
+
{\frac{1}{{\rm Re}_m}}\int |\nabla {\bf B}|^2 d^3x,
\label{D_calE_definition}
\eeq
where we have used the notation
$|\nabla {\bf U}|^2 \equiv
\sum_i |\nabla U_i|^2$ for $i=x,y,z$ and similarly
for ${\bf B}$.
Clearly, the surface term ${\cal S}$ (\ref{S_calE_definition}) includes possible
energy injection into the domain by influx of matter as well as through viscous and resistive stresses.
The term ${\cal D}$ (\ref{D_calE_definition}) represents bulk dissipation.

\subsection{Numerical experiments}

We have performed a number of numerical experiments with the purpose of examining
the effects of the kind of BC employed (periodic versus wall) and of the SB size
(relatively to a relevant integral scale $\ell$ of the problem) on the evolution
of the SB energy content. We have chosen to focus on the total energy ${\mathfrak E}$ of
the domain, as this quantity has (at least to us) the most definite and best understood
physical meaning.

The essence of the problem, as we see it, is embodied in the BC applied in
the shear-wise ($x$) direction and therefore we opt for simulating the evolution
of the SB dynamical equations, restricted to two dimensions -  $x$ and $y$. The
vertical dimension, as important as it may be (we have discussed, for example,
the problems that arise from periodic-BC in $z$) should have no bearing on the
effects we wish to demonstrate here. Using a 2-D code makes the calculation tractable,
but, as is well-known, one cannot expect a persistent activity in two dimensions
as any 2-D turbulence ultimately decays if it is not maintained by some kind
of forcing. In a 3-D SB simulation the activity need not a priori decay, but this
fact should not be important for our basic findings, which comprise of effects
occurring on a short (as compared to the viscous decay) time scale. We shall further
comment on this matter below, after describing our calculation and its results.

We shall consider the evolution of the total domain energy
for two sets of boundary conditions. The first of these are the \underline{periodic-BC},
as laid out before in the paper.  The second type of BC are
appropriate for a rotating channel and we refer to them as \underline{wall-BC}
They implement at the shear-wise boundaries i) no normal flow;
ii) no normal magnetic field flux; and iii) free-slip boundary conditions;
i.e.,
\beq
u = 0, \ b_x = 0, \ \partial_x v_y = 0, \ \ {\rm at} \ \ x = \pm L_x/2.
\eeq
Wall-BC are periodic in the azimuthal direction.

Note that the {\em equations} solved are identical for both sets of BC - they
are the two-dimensional restriction of the SSB equations (\ref{ssdiv})-(\ref{ssinduction}).
We take $\beta=1$ and remark that since there is no $z$ dependence in our 2-D setup,
$B_0$ (the background vertical field) does not play any dynamical role in the
problem. In addition, it is convenient in this case to use
the flux function $\Phi$ and the stream function $\psi$, such
that
\beq
\partial_x \psi = v, \ \  \partial_y\psi = -u, \qquad {\rm and} \qquad
\partial_x \Phi = b_y, \ \ \partial_y\Phi = -b_x,
\eeq
and so the solenoidal property of the velocity and magnetic field disturbances
is automatically insured. It is perhaps instructive to also remark that
the energy integral (\ref{total_E_int}) appropriate for our 2-D numerical calculation
takes on the form
\beq
\dot\mathfrak{E} =
q\Omega_0\int \Bigl[uv - \frac{1}{\beta}b_x b_y\Bigr]^{L_x/2}_{-L_x/2}dy,
 - {\cal D}. \label{totalE_2D_PBC}
\eeq
for periodic-BC, and
\beq
\dot\mathfrak{E} =
\frac{1}{{\rm Re}}\int \Bigl[v(-q\Omega_0)\Bigr]^{L_x/2}_{-L_x/2}dy
 - {\cal D}
 \label{totalE_2D_wall}
\eeq
for wall-BC. Thus, as is apparent, the surface term $\cal S$ reduces
to a simple and readily understandable form.

The code we use in the experiment with periodic-BC is a modification
(so as to include the MHD terms) of the spectral
evolver we developed for the purely  hydrodynamic problem,
see Umurhan \& Regev (2004) for details.  The numerical
experiment employing wall-BC also consists of a spectral code, but
there a different
spectral function basis is used, appropriate for wall-BC.
In both cases, we evolve the flux function $\Phi$ and the stream function $\psi$,
instead of $u$, $v$, $b_x$ and $b_y$, as explained above.
We temporally evolve the resulting set of equations using
the modified Crank-Nicholson scheme as implemented in Umurhan \& Regev (2004).
We investigate the dynamics of a flow with Re = 2000 and Rm = 700 and
make sure that the resolved dynamical scales are in the viscous
and resistive regimes so as to guarantee that the dissipation is
fully resolved in all simulations.
For both types of BC we run the simulations for a SB of size $L_x=\pi$
and $L_y= 2 \pi$ as well as for an SB of double size (in $L_x$, that is $L_x=2 \pi$).

The resolution in the x-direction in the small domain ($L_x = \pi$ and $L_y = 2\pi$)
is taken to be $n_x = 64$ and $n_y = 64$. Because of our experience with externally
driven shear problems (e.g. Umurhan \& Regev, 2004), we are careful to
to increase the resolution in the shear-wise direction
on account of the strong crenellation that the external
shear creates out of the disturbances (the Orr-mechanism), so that
for the double (in the shear-wise
extent of the domain) case, we double the resolution too, i.e.,
when $L_x = 2\pi$ we take $n_x = 128$.

\begin{figure}
\begin{center}
\leavevmode \epsfysize=6.cm
\epsfbox{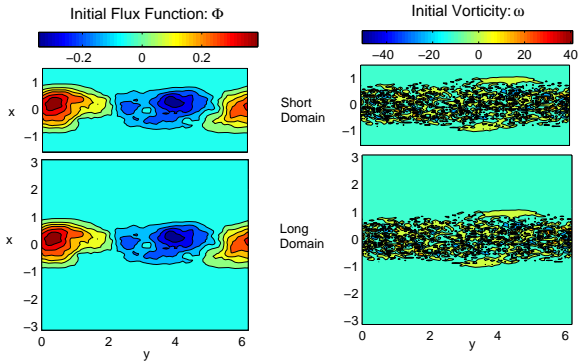}
\end{center}
\caption{ The initial conditions used
for the simulations.  The perturbations are localized
away from the radial boundaries and are identical for
the periodic-BC and wall-BC runs.
} \label{Initial_Conditions}
\end{figure}

The simulations are initiated with white noise in
the vorticity ($\omega=\nabla^2 \psi$) and the source $J$
of the flux function ($J=\nabla^2 \Phi$). The initial
$\omega$ and $\Phi$ are shown in
Fig. \ref{Initial_Conditions}.
This initial disturbance is identical in all simulations
and it is localized away from the radial boundaries.
We do this to gain better control of the effects that the boundaries
have in the evolution. The only relevant spatial dynamical scale that may be perceived
as the integral, or injection, scale is the extent of the initial perturbation,
suggesting that $\ell \sim 2$, initially.

\begin{figure}
\begin{center}
\leavevmode \epsfysize=7.5cm
\epsfbox{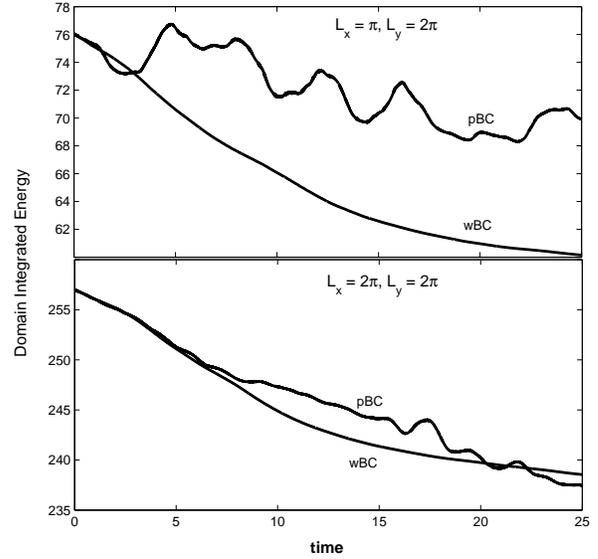}
\end{center}
\caption{  The total energy (in arbitrary
units) in the SB as a function of time (in units of $\Omega_0^{-1}$).
In the upper panel, the size of the SB is $L_x=\pi$ and $L_y= 2 \pi$
and the result of the calculation with periodic-BC is compared
with the one using wall-BC. In the lower panel, the same is shown
for a SB having a double shear-wise extent.
In both cases the simulations are started with the same initial conditions,
as depicted in Fig. \ref{Initial_Conditions} and the evolution is shown
during the period in which the typical (of 2-D) decay of the activity
has not yet fully set in.}
\label{Energy1}
\end{figure}

In the upper panel of Fig. \ref{Energy1}, we show the evolution of the total energy in
the SB of shear-wise extent $L_x = \pi$, for periodic-BC and wall-BC, while this figure's
lower panel depicts the same for a SB of double shear-wise extent ($L_x = 2 \pi$).
The fluctuations in $\mathfrak{E}$ depicted on the upper panel
in periodic-BC simulations are quite dramatic and we have found that they correlate with the
action taking place on the boundaries. Clearly, in the smaller SB we have
$L_x \sim \ell$, that is we are in the situation in which the periodic box size
is too small, as we have discussed at the end of the previous section. The dynamics,
for the periodic-BC case, is seriously affected by the artificial distortion
of the correlation function brought about by the periodic-BC on a too small domain.
In contrast, the evolution with wall-BC proceeds smoothly, as the boundary conditions
do not allow artificial energy inflow/outflow.
The fluctuations in the periodic-BC case have a time scale
related to the period time of the shearing box given by
\[
T = \frac{L_y}{\Omega_0q L_x},
\]
as we can clearly see from the time behavior of the surface term
(not shown). This behavior is also apparent in Fig. \ref{Energy1},
but is less prominent and clear.
We reason that these fluctuations are driven by successive passages of
the developed coherent structures (vortices, in this case) in the
imaginary neighboring boxes. The vortices that appear in all our simulations
are actually somewhat tighter than the initial vorticity perturbation extent,
suggesting that $\ell \sim 1$ or so.
We also see this trend played out in the lower panel, however an important difference
is clearly apparent. The domain size is now $L_x \ga 6$, that is, considerably
larger than $\ell$. In this case, the periodic-BC do not induce very significant
spurious energy exchange with the exterior and consequently the fluctuations
of the SB energy content are rather mild. We may reasonably conjecture that
a calculation with $L_x\gg \ell$ would reveal no difference between the pBC and wBC
curves.

The longtime decay, which we have mentioned above as being typical in a 2-D problem
of this kind, is obviously due to the bulk dissipation embodied in the term $\cal D$.
We comment again that the effects we wished to demonstrate in the above numerical
experiment can not be significantly altered by the third (vertical) direction
because the key factor responsible for them is the boundary condition in the
{\em shear-wise} direction.

In summary, our numerical experiments show that the type of boundary
conditions, coupled with the box size, adopted in investigations of local
disk turbulence have important implications on the physical results.
When one considers the total energy of disturbances within
the formalism of the SB equations, we note that the total
energy $\mathfrak{E}$ fluctuates
dramatically on account of the injection/extraction of
energy across the boundaries when the size of the box is not
large enough as compared to the relevant injection scale.
It is thus impossible to see how a SB simulation with periodic-BC can reveal
any dynamical behavior that is relevant to a true accretion disk,
in which $\ell$ has to be at least of the disk thickness size.

\section{Discussion and Summary}
\label{final}

Based on the arguments of the previous sections we are led to conclude that
existing SB calculations, done with periodic-BC in the fixed-flux as well as zero-flux cases,
suffer from some serious inconsistencies and problems.
These inconsistencies make these calculations quite unreliable
for their purpose - finding the mechanism of and quantitatively
estimating the turbulent angular momentum transport and energy
dissipation in accretion disks. As we have discussed
in this paper, the problems stem from the limitations of the box scale,
insufficient resolution of the numerics, unjustified symmetry properties
(due to periodic-BC) and, as an additional consequence of periodic-BC
and limited box size, the possibility of unrealistic energy sources (sinks),
which can seriously interfere with the dynamics.

In the fixed-flux case the linear MRI gives rise
to a supercritical transition. In this case
the SB calculations with periodic-BC manifest channel flows, resulting from the nonlinear
exponential growth of the CM, whose problematic nature arises not only
from their shear-wise symmetry but also
the fact that the linear instability appears to be more involved than just
the one comprising of a linear superposition of the CM.
The zero-flux simulations in the same SB setup do not
allow for CM, but they suffer from all the other difficulties.
Indeed, also zero-flux numerical calculations, when tested for convergence
(most recently those of Fromang \& Papaloizou 2007), indicate that in the
ideal case $\alpha$ (a measure for the angular momentum
transport) decreases steadily with increasing resolution.
Moreover, in the zero-flux case a dynamo action must be invoked
and the MRI is thought to play a role in it. But it is not clear how
such an intricate self-sustained process, in which a subcritical
transition, similar to the one occurring in hydrodynamic shear flows
(see ROP07), can be faithfully uncovered by SB periodic-BC simulations,
with their problems, as discussed in this paper.

KPL07 conjectured that the fixed-flux case is probably irrelevant to
accretion disks, but clearly accretion disks exist
that are threaded by external fields (like, e.g., those around young stellar
object, see Shu {\em et} al. 2007). Moreover, Pessah, Chan \& Psaltis (2007)
claim the opposite, i.e. that MRI related angular momentum transport
can only occur in disks threaded by strong ($\sim$ equipartition value) fields, basing
this conclusion on their numerical simulations. In any case,
even if the ultimate development of the magnetic
field in the disk makes its geometry inside very involved, this fact cannot annihilate
the external field whose lines must close through the disk. It seems, therefore,
more likely that the apparent almost linear dependence of $\alpha$ on $B_z$
in some SB fixed-flux simulations is simply due to the non-reliability of these
calculations. Moreover, SB zero-flux calculations also suffer from most of the same
difficulties as the fixed-flux ones (as discussed above).

The most recent numerical studies using the SB with periodic-BC in both fixed-flux and zero-flux case
have incorporated explicit $\rm Re$ and ${\rm Re}_m$ (that is, physical non-zero viscosity
and resistivity) and studied the dependence of the angular momentum transport (or $\alpha$) on
these non-dimensional numbers. As we have indicated before, this approach
can yield valuable information.
Lesur \& Longaretti (2007) and Fromang {\em et} al. (2007) have reported
that $\alpha \propto {\rm P}_m^{\kappa}$ with $\kappa>0$.
We have shown earlier, using asymptotic semi-analytical techniques,
(Umurhan, Menou \& Regev, 2007) that such scaling can be naturally expected
for ${\rm P}_m \ll 1$, and also found that $\kappa$ depends on the boundary conditions
(URM07). Our work was done for fixed-flux conditions and in a thin-gap mTC setup
(which eliminates the shear-wise invariant CM).
However, in general, any dependence of the angular momentum transport
on a positive power of ${\rm P}_{m}$ makes the role that the MRI takes
as the driver of angular momentum transport in astrophysical disks
with such astronomical Reynolds numbers rather doubtful
(at least in the fixed-flux case, where it is considered to be the sole driver).

Laboratory experiments have so far not yielded any conclusive
indications of MRI driven turbulence, save for the case
in which relatively strong azimuthal fields are applied
(Hollerbach \& R\"udiger 2005; Stefani {\em et} al. 2006; Szklarski \& R\"udiger
2007). In thin accretion disks, however, strong toroidal (or any other, for that
matter) magnetic fields cause significant magnetic pressure and are likely to
be inappropriate for accretion disks in the thin, rotationally supported,
disk limit.

It is not easy to answer the question, what kind of numerical scheme
or which setup is best suited for the problem?
In any case, it is not clear what is less appropriate for
local numerical studies of accretion disks: wall-BC or periodic-BC in
the SB.  We believe that the use of periodic-BC, especially in the
radial direction, suffers from serious problems unless
the power in the perturbation fields is localized away from the
boundaries of the system (Dubois {\em et} al. 1999, Davidson 2004).

We think that if one wants to extract reliable
physical information on the local nature of MHD turbulence
in which shear is instrumental, on scales from the dissipation one
and up to well into the inertial
range, it is likely that
mrpC (e.g., ROP07) or mTC (e.g., Liu, Goodman \& Ji 2006; Obabko, Cattaneo
\& Fischer 2006) in a thin-gap setup are
more appropriate. There certainly are no walls in accretion disks,
but wall-BC calculations of this kind can, at least, be made mathematically
consistent and physically sound.
One may
think that periodic-BC (which are also {\em not} realistic for accretion disks) are better
suited, but as we have shown SB calculations with periodic-BC suffer from
difficulties and inconsistencies for this problem, mathematical as well as physical.

It seems to us that the logical approach to the accretion disk angular momentum
transport problem should thus consist of:
\begin{enumerate}
\item
mathematically and physically sound calculations, most likely
in the mrpC or mTC thin-gap setups, for the purpose of learning about
the detailed local properties of MHD turbulence
(its energy spectrum, correlations etc.);
\item
devising a turbulence model, based on the above, so as to replace
the $\alpha$ model;
\item
global calculations employing the above turbulence model(s) and
various boundary conditions, giving rise to accretion disk
models that can be confronted with observations.
\end{enumerate}
Regarding the first item, it has to be kept in mind that the saturation
mechanism may perhaps be different from the case when there are no physical
boundaries (see Knobloch \& Julien 2005, who have performed
an asymptotic MRI analysis, but for a developed state, far from marginality).
The hope is, however, that this should not affect too much the local properties
of the MHD turbulence.

The previously mentioned approach of performing global calculations
with low Reynolds numbers (e.g., K\'ersale {\em et} al. 2005) are
also useful as complementary studies, in which the dependence of
the results on relevant non-dimensional numbers can be uncovered.
In any case, any numerical study must be controlled, in the sense
that one can faithfully (that is, in a numerically resolved
way and without introducing any a-priori periodicity
of disturbance correlations) experiment with the relevant mathematical
properties (e.g., BC, symmetries) and physical quantities
(e.g., viscosity, resistivity) and uncover their effects.
Schemes including {\em hyper-viscosity}, which are routinely used
in numerical turbulence studies, may be employed, but this
non-trivial tool must be used with proper care.
Global quantities (like the energy) must obviously also be monitored
and their development fully understood.

The problem of angular momentum transport in accretion disks, especially
if MHD turbulence in a non-isotropic strongly sheared medium is the
physical mechanism responsible for it, is extremely difficult and
involved. This work, as well as that of KPL07 and others, indicates that
the understanding gained so far from SB and other numerical simulations
seems unfortunately not to be adequate enough to provide a viable
physical prescription that can be effectively used in modeling accretion disks,
beyond the old $\alpha$-viscosity parametrization.
It seems that only an extensive and coherent program,
of the kind that has been undertaken in the turbulent dynamo problem
(see Schekochihin {\em et} al. 2007), is likely to lead to significant progress.
In such a program a massive numerical effort is indispensable, but
we would like to stress, in concluding, that analytical (asymptotic or other)
approaches can also be very useful in guiding the numerical and laboratory work
and better understanding the underlying physics.

\section{Acknowledgements}

We thank Kristen Menou for his comments and suggestions and Amos Ori for a useful discussion.
We gratefully acknowledge Bruno Coppi for pointing out to us his relevant work and for
sharing with us some of his deep understanding and insight on the subjects of this paper.
Michael Mond shared with us the results of his research, which is in progress, read
the manuscript and commented on it. We are grateful to him for this. Finally, we thank Fausto Cattaneo and Aleksandr Obabko
who kindly showed us, prior to publication, the results of their extensive high-resolution simulations
and discussed them with us.

\end{document}